\documentstyle[epsf]{mn}
\def\dgr{^\circ}
\newcommand{\etal}{{et al}\/.}
\newcommand{\source}{PKS 0521$-$365}
\begin{document}
\title[Extended X-ray emission from \source]{Extended X-ray emission from the BL Lac object \source}
\author[M.J.~Hardcastle \etal]{M.J\ Hardcastle$^1$, D.M.\ Worrall$^{1,2}$ and M.\
Birkinshaw$^{1,2}$\\
$^1$Department of Physics, University of Bristol, Tyndall Avenue,
Bristol BS8 1TL\\
$^2$Harvard-Smithsonian Center for Astrophysics, 60 Garden Street,
Cambridge, MA 02138, U.S.A.}
\maketitle
\begin{abstract}
Models that seek to unify BL Lacs and low-power radio galaxies predict
that the two types of object should show similar isotropically emitted
X-ray emission. Testing this is usually limited by difficulties in
separating strong X-ray emission from a BL Lac nucleus and surrounding
low-surface brightness emission.  In this paper we report ROSAT HRI
observations of the $z=0.055$ BL Lac object \source . We are able to
separate a luminous extended X-ray component from the bright
nucleus. Using a new radio map, we show that it is unlikely that the
extended emission is due to inverse-Compton scattering of photons from
the active nucleus, and instead interpret it as thermal emission from
dense, rapidly cooling gas. This is a more extreme environment than is
found in typical FRI radio galaxies, and may pose a problem for
unified models.

\end{abstract}

 
\section{Introduction}

In unified models for radio sources, BL Lac objects are the
low-power radio galaxies whose jets are the most highly beamed along our
line of sight (e.g.\ Browne 1983; Antonucci \& Ulvestad 1985; Ulrich
1989; Urry \& Padovani 1995). Their rapid variability, apparent
superluminal motion, strong point-like emission in radio, optical
and X-ray, and the detection of some sources in $\gamma$-rays, are all
explained if we are seeing emission from a relativistic jet closely
aligned with our line of sight. Low-power radio galaxies then represent the
`parent population' of unaligned objects whose jets are less
favourably aligned. These low-power radio galaxies are likely mostly
to be Fanaroff \& Riley (1974) class I (FRI) objects, but the parent
population may also include some transitional objects and
low-excitation FRII radio galaxies.

An important test of this unified model is the degree to which the
isotropic (alignment-independent) properties of BL Lacs are similar to
those of the parent population of radio galaxies. Such tests have been
made, on the whole successfully, by looking at the extended radio
emission (e.g.\ Antonucci \& Ulvestad 1985; Kollgaard \etal\ 1992;
Perlman \& Stocke 1993, 1994) and properties of the host galaxies
(e.g.\ Ulrich 1989; Abraham, McHardy \& Crawford 1991; Wurtz, Stocke
\& Yee 1996; Falomo 1996) although there is some evidence that there
are too few BL Lacs associated with the most luminous host galaxies
(Wurtz \etal\ 1996).

Another isotropic indicator is the clustering environment. Using
two-point correlation analysis of optical fields, it has been shown
that FRI radio galaxies are normally found in groups or clusters of
galaxies (Longair \& Seldner 1979; Prestage \& Peacock 1988) and
BL~Lacs seem also to inhabit groups or poor clusters (Pesce, Falomo \&
Treves 1995; Smith, O'Dea \& Baum 1995; Wurtz, Stocke \& Ellingson
1997) though it appears that, at least at low redshift, BL Lacs
are not often found in the dominant galaxies of rich clusters
(Prestage \& Peacock 1988; Owen, Ledlow \& Keel 1996; Wurtz \etal\
1997); for example, Owen \etal\ (1996) find no BL Lacs at the centres
of Abell clusters, a result inconsistent at the 95\% confidence level
with the numbers expected from the unified models of Urry \& Padovani.

Clustering environment may also be investigated by X-ray
observations. It has long been known that many objects close to the
FRI-FRII luminosity boundary are associated with rich clusters having
luminous X-ray haloes. Recent observations with {\it ROSAT} have shown
that more typical FRI radio galaxies have extended thermal X-ray
emission on scales characteristic of groups or poor clusters (Worrall
\& Birkinshaw 1994). This offers a new way to test the unification
hypothesis; such emission should be isotropic, and so we expect all BL
Lacs to have X-ray haloes comparable to those of FRIs. This test is
difficult because it requires us to separate any extended X-ray
emission from the bright unresolved emission of the BL Lac nucleus. In
this paper we describe such an analysis of {\it ROSAT} observations of
the BL Lac \source .

\source\ is a well-studied BL Lac with a redshift of 0.055, comparable to the
redshifts of the radio galaxies studied by Worrall \& Birkinshaw (1994). It
is variously described in the literature as a blazar, a BL Lac object,
or an N-galaxy, and on multifrequency spectral index plots like those of
Sambruna, Maraschi \& Urry (1996) is placed among radio-selected BL
Lacs. Its host galaxy is easily detectable in the optical
[contributing $\sim 50$ per cent of the source luminosity at 5500 \AA\
in an $8 \times 8$ arcsec effective aperture; Falomo, Scarpa \&
Bersanelli (1994)] and it exhibits strong, variable broad emission
lines (Scarpa, Falomo \& Pian 1995). The host galaxy is a giant
elliptical (Wurtz \etal\ 1996). Pesce \etal\ (1995) suggest that the
excess galaxy count around the object corresponds to a cluster of
Abell richness 0 or more; they identify at least one, and up to four
associated galaxies within 100 kpc. However, the cross-correlation
analysis of Wurtz \etal\ (1997) suggests a poorer cluster, with
richness class $<0$.

\label{unbeamed}
In the radio, the source has a 408-MHz flux of 36.1 Jy (Wright \&
Otrupcek 1990), corresponding to a power at that frequency of $\sim 4
\times 10^{25}$ W Hz$^{-1}$ sr$^{-1}$; this places it slightly above
the nominal FRI-FRII luminosity boundary ($\sim 1 \times 10^{25}$ W
Hz$^{-1}$ sr$^{-1}$ at this frequency), though of course some of the
408-MHz emission is contributed by the core, presumed to be beamed. It
exhibits a core-halo-hotspot morphology on arcsecond scales (Wardle,
Moore \& Angel 1984; Ekers \etal\ 1989; see also section
\ref{discuss}), which, together with its comparatively high radio
power, may suggest that it is an aligned version of a transitional
FRI-FRII object. The prominent radio jet is also seen in optical
synchrotron emission, extending about 6 arcsec from the nucleus (e.g.\
Keel 1986, Macchetto \etal\ 1991). No motion of the core components was
detected in VLBI observations (Tingay \etal\ 1996) and this, together
with the comparatively low ratios of nuclear to host-galaxy optical
emission (Falomo \etal ) and radio core to extended radio flux
(Antonucci \& Ulvestad 1985), suggests a source that is only
moderately relativistically boosted along the line of sight compared
to the most extreme BL Lacs. It was for this reason that we selected
it as a suitable candidate for an X-ray search for extended emission
with the {\it ROSAT} HRI. \source\ has already been extensively
observed at X-ray wavebands, with {\it Einstein} (Worrall \& Wilkes
1990), {\it EXOSAT} (Sambruna \etal\ 1994) and the {\it ROSAT} PSPC
(Pian \etal\ 1996), and was detected in $\gamma$-rays by {\it EGRET}
(Lin \etal\ 1996), but none of these observations provides the
resolution necessary to separate compact from extended structure. The
source appears to have a concave spectrum in the
X-ray waveband; the energy index is $\sim 1.0$ in the soft band,
flattening to $\sim 0.5$ at higher energies.

Throughout the paper we use a cosmology in which $H_0 = 50$ km s$^{-1}$
Mpc$^{-1}$, $q_0 = 0$. At the distance of \source\, 1 arcsec
corresponds to 1.475 kpc.

\section{Observations}

We observed \source\ with the {\it ROSAT} HRI for 46.6 ks between 1995
Feb 09 and 1995 Feb 11. The data were analysed using the IRAF Post-Reduction
Off-line Software (PROS).

\source\ is detected at $10620 \pm 140$ net counts in a 2.5-arcmin radius
circle with background taken from an annulus between 2.5 and 3.5
arcmin. The source circle was chosen by expanding the radius until the
background-subtracted counts enclosed stayed constant within the errors. A
25-arcsecond radius circle around a nearby point source (1 arcmin to
the north) was excluded from the analysis. The detected counts
correspond to a rate of $0.228 \pm 0.003$ counts s$^{-1}$. The source
centroid is consistent to within a few arcseconds with the best
available optical and radio positions.

\label{var}
A simple analysis, binning the data into 1-ks intervals, suggests that
the source was not strongly variable during the 110 ks that contain
our observations. To compare the source flux with the earlier {\it
ROSAT} PSPC observations, we re-analysed the 4.8 ks of PSPC data from
the {\it ROSAT} public archive, taking account of the nearby
background sources. We detect the source at $3610 \pm 60$ PSPC counts
in the same source region, or $0.75 \pm 0.01$ counts s$^{-1}$. A
spectral fit to a power-law model between 0.1 and 2.4 keV using the
galactic $N_H$ ($3.37 \times 10^{20}$ cm$^{-2}$) of Elvis, Lockman \&
Wilkes (1989) gives an energy spectral index of $0.925\pm 0.03$
($\chi^2 = 23$ with 32 d.o.f.; errors are $1\sigma$ for one
interesting parameter), consistent with the results of Pian \etal\
(1996) and Sambruna (1997); the unabsorbed 1-keV flux density from the
PSPC data is then $2.03 \pm 0.03$ $\mu$Jy, while the flux density from
the HRI data on the same assumptions is $1.88 \pm 0.02$ $\mu$Jy. If
$N_H$ is allowed to vary the best-fit spectral index is $1.10\pm 0.12$
with $N_H = 4.0_{-0.4}^{+0.3} \times 10^{20}$ cm$^{-2}$ ($\chi^2 =
17.8$ with 31 d.o.f.; errors $1\sigma$ for two interesting
parameters). Incorporating the spectral uncertainties, the flux in the
0.1--2.4 keV {\it ROSAT} band seems to be $10 \pm 2$ per cent lower in
the HRI data than it was at the epoch of the PSPC observations (1992
Aug 29). We return to this point below.

\section{Analysis}
\label{analysis}

\begin{figure*}
\begin{center}
\leavevmode
\epsfxsize 8cm
\epsfbox{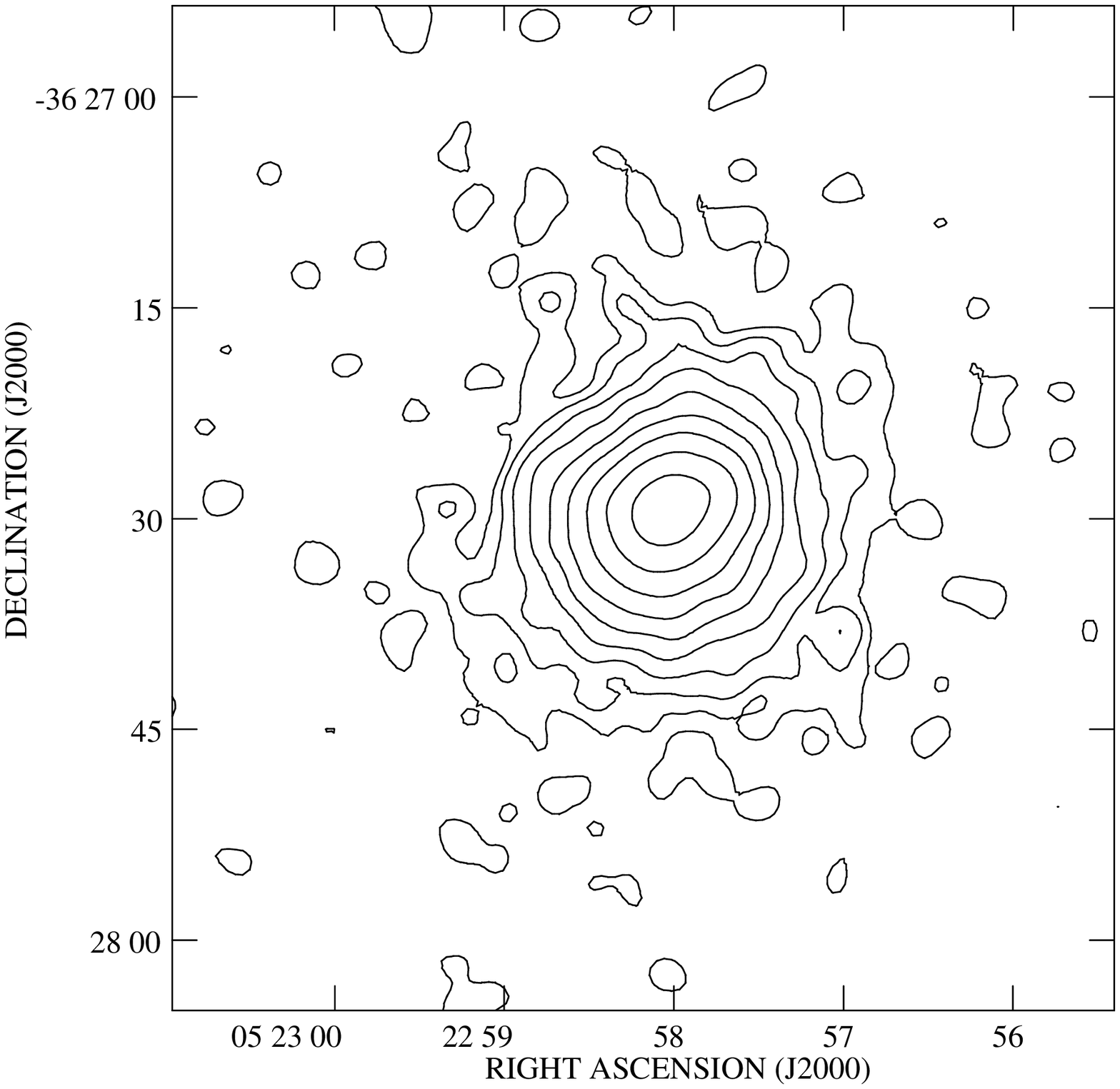}
\epsfxsize 8cm
\epsfbox{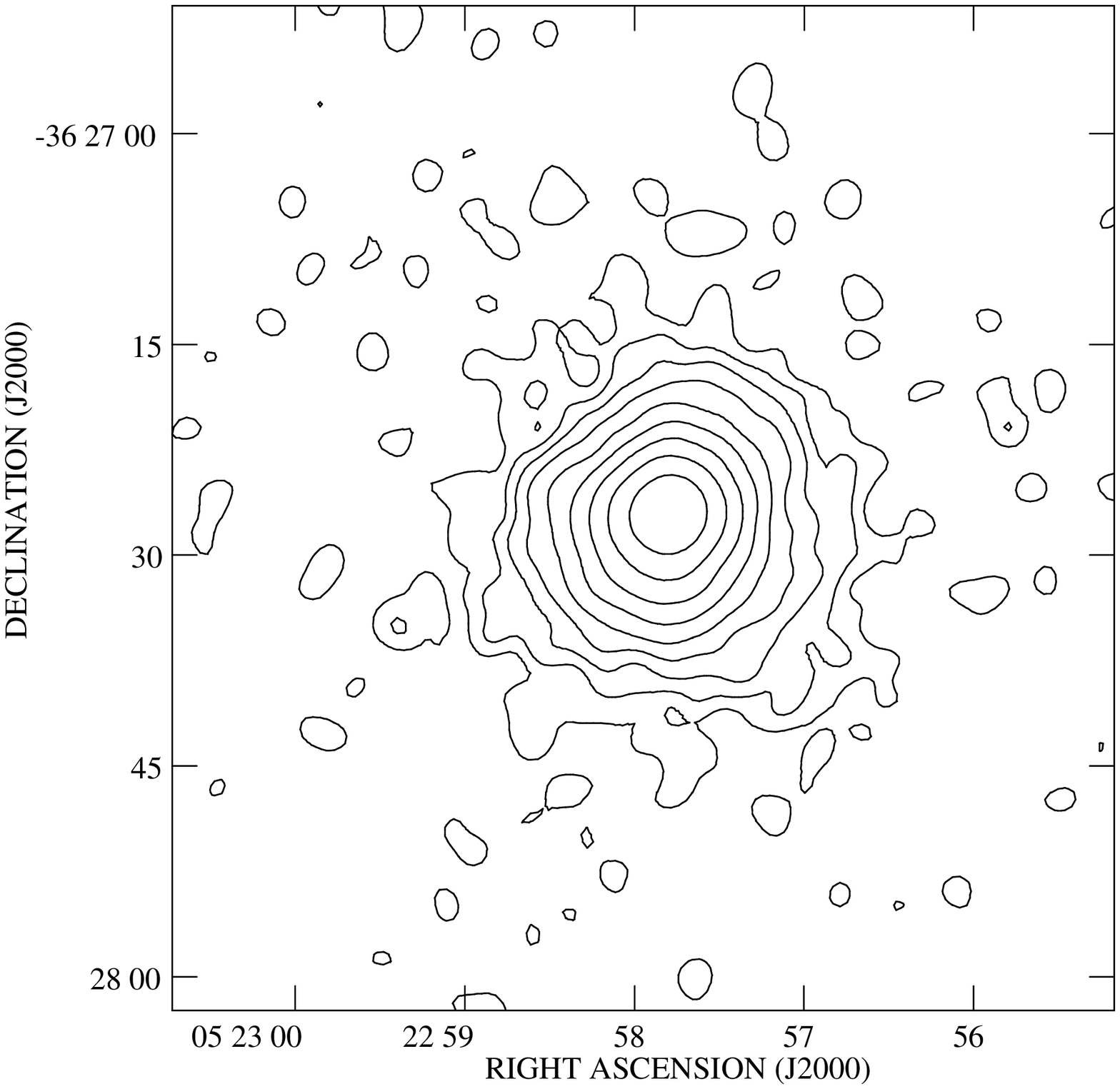}
\caption{Left: contour plot of the {\it ROSAT} HRI data for the inner
regions of \source , smoothed with a Gaussian of $\sigma = 1$
arcsec. Contours at $0.4 \times (1, 2, 4, \dots)$ counts
arcsec$^{-2}$; the lowest contour corresponds approximately to the
$3\sigma$ (99.87 per cent confidence) level for this convolving
Gaussian (see Hardcastle, Worrall \& Birkinshaw 1998c). Right: as
left, but showing the data after the dewobbling procedure had been
applied.}
\label{contour}
\label{rest}
\end{center}
\end{figure*}

\begin{figure*}
\begin{center}
\leavevmode
\epsfxsize 8cm
\epsfbox{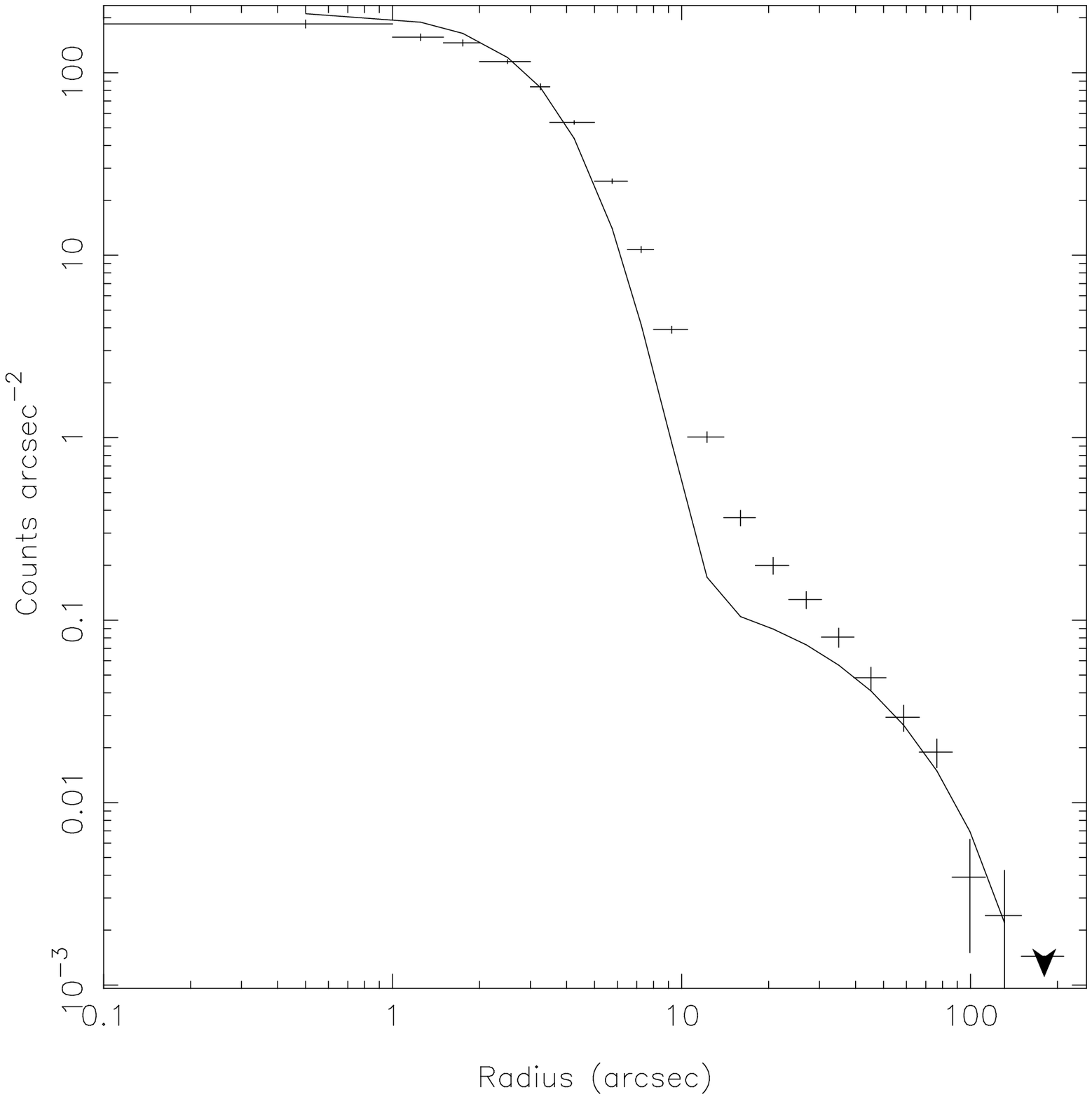}
\epsfxsize 8cm
\epsfbox{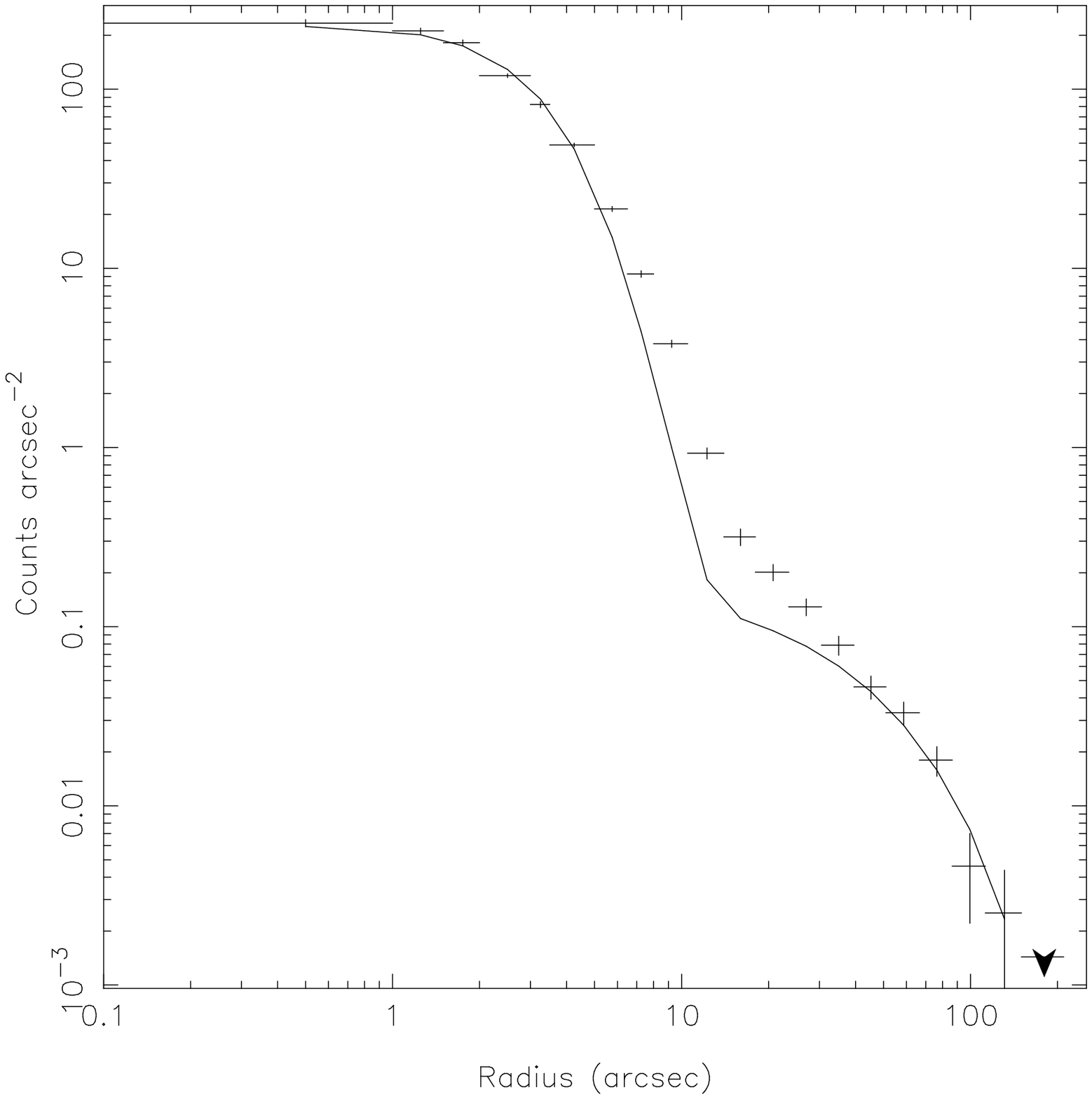}
\caption{Radial profile of \source ; counts arcsec$^{-2}$ as a
function of distance from the source centre. The nominal on-axis PRF of the
ROSAT HRI, normalised to the total counts within 5 arcsec of the
centre, is plotted for comparison (solid line). Left: before
dewobbling procedure. Right: after dewobbling procedure.}
\label{radial}
\label{rradial}
\end{center}
\end{figure*}

Fig.\ \ref{contour} (left) shows that the source is approximately
symmetrical; radial profiling is therefore appropriate. To search for
extension in \source\, we first compared its radial profile with the
standard empirical expression (David \etal\ 1997) for the HRI point
response function (PRF).  It will be seen from Fig.\ \ref{radial}
(left) that the radial profile is not well fitted by a standard {\it
ROSAT} PRF (normalised to the central regions, which we expect to be
dominated by the point source); there is a deficit of counts at small
radii and an excess of counts at large ones.

It is well known that images from the HRI can be badly affected by
errors in the {\it ROSAT} aspect correction; the effect is to produce
spurious elongation (`smearing') in the images of point sources, thus
broadening the effective, azimuthally-averaged PRF. Such smearing is
likely to be significant compared to the detector and instrument point
response on scales of $\la 10$ arcsec. As shown in Fig.\ \ref{contour},
the source appears elongated, in its central regions,
in position angle roughly $135\dgr$; aspect smearing could therefore be
responsible for some of the deviations from the nominal PRF shown in
Fig.\ \ref{radial}.

The aspect errors are thought to arise because the star tracking
system does not always calculate positions correctly, due to
variations in the gains of the pixels of the star tracker CCD. The
spacecraft wobble exacerbates this problem because it causes the guide
stars to move across the tracking CCD on short timescales. Methods for
correcting aspect smearing (e.g.\ Morse 1994) therefore rely on
binning the data according to the phase of the wobble; so long as the
satellite roll angle and the properties of the CCD are constant, the
aspect error should be a function of the wobble phase only. To do
this, we applied {\sc iraf}/PROS scripts provided by J.\ Silverman and
D.\ Harris, based on suggestions by G.\ Hasinger (for more details see
Harris \etal\ 1998). We first established that all our data were
taken at the same roll angle; this, along with the relatively small
number of closely spaced observation intervals, allowed us to analyse
all the data together. The scripts bin the data as a function of
wobble phase, utilising the fact that the relative wobble phase at any
time is given simply by the spacecraft clock (modulo clock resets,
none of which occured during our observation). We chose to divide the
data into 20 phase bins, giving approximately 500 counts per binned
observation; with 20 phase bins there are enough counts for
centroiding in each bin, and examination of images derived from each
bin showed them to be circular. A centroid was then found, using the
{\it detect} suite of tasks within PROS, for each individual bin, and
the images were restacked so that the centroids aligned.

The resulting image (Fig.\ \ref{contour}, right) appears less
elliptical than the uncorrected image (Fig.\ \ref{contour}, left) and
is considerably better fit, in its inner regions, by the nominal HRI
PRF (Figure \ref{radial}, right); it appears that the stacking
procedure has made the core slightly narrower than the nominal PRF,
which is of course derived from observations that may themselves have
been affected by aspect uncertainties. We conclude that the inner
regions of \source\ are well described as a point source. More
importantly, even after dewobbling we still see an excess of counts
over the PRF at radii of 10-40 arcsec. There is no reason to doubt
that this corresponds to real extended emission.

\begin{figure*}
\begin{center}
\leavevmode
\epsfxsize 11 cm
\epsfbox{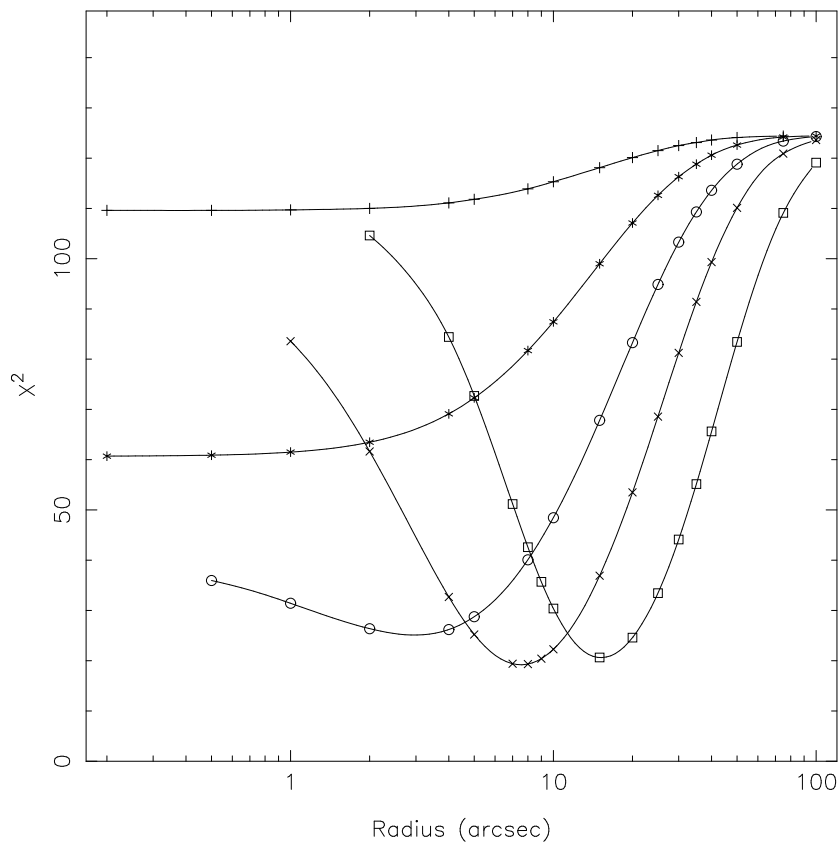}
\caption{$\chi^2$ as a function of core radius of the $\beta$ model
for fits of the radial profile of \source\ to a model consisting of
the sum of a $\beta$ model and point-like emission, both convolved
with an HRI PRF. Points marked are the results of fits
while the lines are natural cubic splines drawn through the points. 16
radial bins were used. The best fit (with $\chi^2 = 19$, 14 d.o.f.)
has $\beta=0.9$, core radius 8 arcsec.  Plusses
indicate fits with $\beta = 0.35$, stars $\beta=0.5$, circles $\beta =
0.67$, Xs $\beta = 0.9$ and squares $\beta=1.5$.}
\label{chi2}
\end{center}
\end{figure*}

\begin{figure*}
\begin{center}
\leavevmode
\epsfxsize 11 cm
\epsfbox{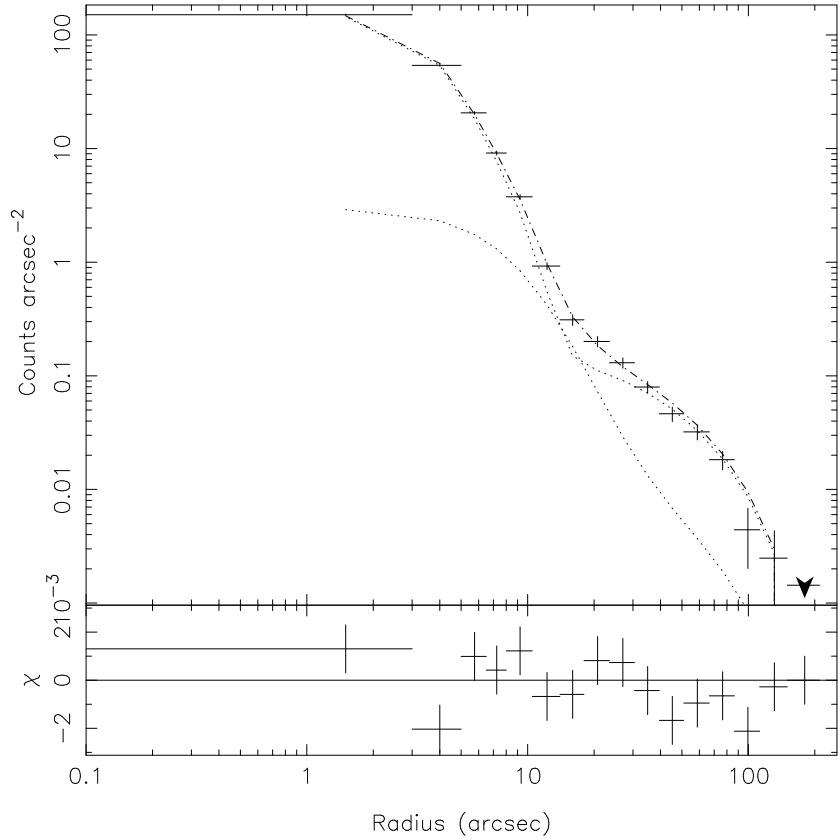}
\caption{The radial profile of \source\ plotted with the best-fit
combination of $\beta$ model and point source (dot-dash line). The
contributions of the separate model components are shown as dotted
lines. The lower part of the plot shows the residual between data and
model expressed in units of the fitting statistic $\chi$. A background
level of $6.8 \times 10^{-2}$ count arcsec$^{-2}$ has been subtracted.}
\label{radialmod}
\end{center}
\end{figure*}

To characterise the scale size of this extended emission we fit models
consisting of a $\beta$-model (Sarazin 1986) and a point-source
component, both convolved with the nominal PRF, to the
background-subtracted radial profile of the dewobbled image. To avoid
large contributions to the fitting statistic due to the fact that the
restacked data are narrower than the nominal PRF, we sampled the
radial profile more coarsely in its central regions for these fits. We
performed fits for various different values of the parameter $\beta$
and the core radius. Figure \ref{chi2} shows a plot of $\chi^2$ as a
function of $\beta$ and core radius; the best fit ($\chi^2 = 19$ with
15 degrees of freedom) has $\beta = 0.9$ and a core radius of 8
arcsec, with a central normalisation of the $\beta$-model of $5.4 \pm
0.5$ counts arcsec$^{-2}$. Figure \ref{radialmod} shows the best-fit
model plotted with the data. Fitting a point source alone to the data
gives unacceptably high values of the fitting statistic ($\chi^2 =
124$).

\begin{figure*}
\begin{center}
\leavevmode
\epsfxsize 11 cm
\epsfbox{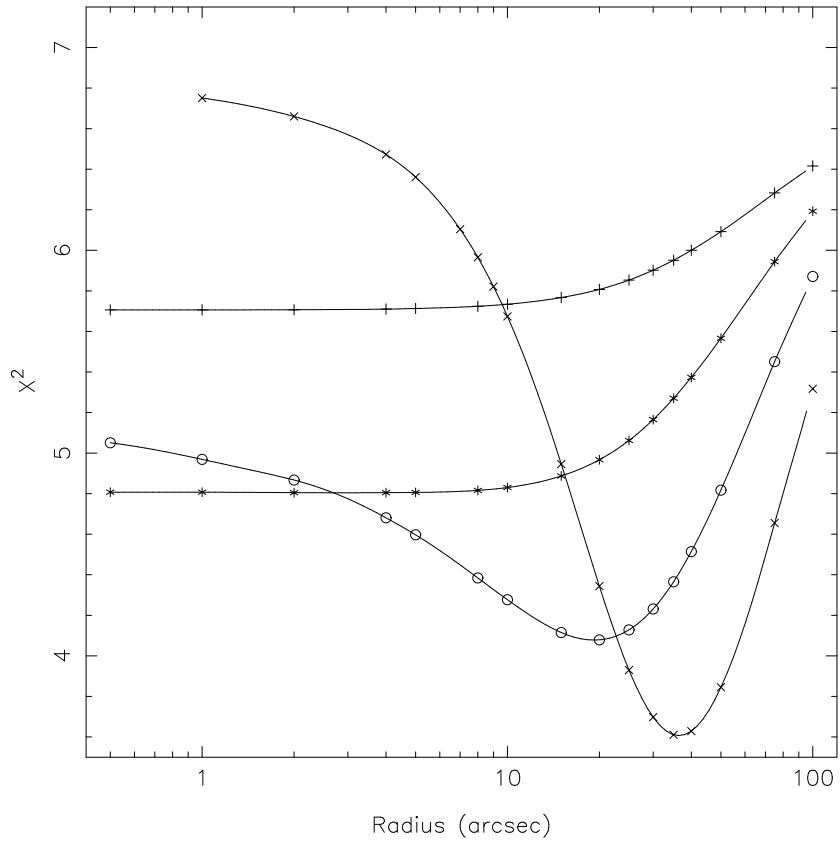}
\caption{$\chi^2$ as a function of core radius of the $\beta$ model
for fits of the PSPC radial profile of \source\ to a model consisting
of the sum of a $\beta$ model and point-like emission. Points marked
are the results of fits while the lines are natural cubic splines
drawn through the points. 12 radial bins were used. The best fit (with
$\chi^2 = 3.6$, 9 d.o.f.) has $\beta = 0.9$, core radius 35 arcsec.  Plusses
indicate fits with $\beta = 0.35$, stars $\beta=0.5$, circles $\beta =
0.67$ and Xs $\beta = 0.9$.}
\label{chi3}
\end{center}
\end{figure*}

For comparison, we fitted a similar range of $\beta$-models to the
archival PSPC dataset. Because of the broader intrinsic PRF of the
PSPC, smearing is not considered to be a problem with this instrument.
The best-fitting point source and $\beta$-model had a core radius of
35 arcsec with $\beta = 0.9$, as shown in Fig.\ \ref{chi3}. The data
are better fit with this model than with a point-source model alone
(best-fit $\chi^2 = 3.6$ with 9 degrees of freedom), although the
point-source fit is formally acceptable ($\chi^2 = 6.9$ with 10
degrees of freedom). A large range of possible core radii are
acceptable, and the HRI best-fitting $\beta$ model is among those
consistent with the PSPC data within the 90 per cent confidence
contour; the normalisations of the models are also consistent. If an
appropriately normalised $\beta$-model with core radius 8 arcsec and
$\beta=0.9$ is subtracted from the PSPC data the remaining emission is
adequately modelled as a point source with no significant support for
an additional extended component. We therefore conclude that the
extended emission is well represented by the HRI best-fit model. The
best-fit point-source count rates in this model in the PSPC and HRI
data are consistent to within a few per cent, suggesting that the
inconsistency between the total PSPC and HRI count rates (section
\ref{var}) is a result of our not having taken into account the two
spectral components corresponding to compact and extended emission.

\section{Discussion}
\label{discuss}

The $\beta$-model fits to the HRI data described above imply that the
extended X-ray emission has a core radius of 12 kpc and contributes
$860 \pm 90$ counts to the total in a 2.5-arcmin radius circle. The
corresponding 0.2-1.9 keV luminosity, assuming a Raymond-Smith model
with $kT = 1$ keV and using galactic $N_H$, is approximately $8 \times
10^{35}$ W. This is considerably more luminous than the extended
emission seen in most of the radio galaxies of Worrall \& Birkinshaw
(1994), although the FWHM of 14 kpc is comparable to some of the
smaller objects. The central cooling time, on the same temperature
assumption, is $6 \times 10^8$ yr, which, if the emission were
thermal, would imply rapid cooling, with expected mass deposition
rates of $\sim 6$ $M_\odot$ yr$^{-1}$. The radius at which the cooling
time is comparable to the Hubble time would be $\sim 20$ kpc, larger
than the core radius, which means that a simple single-temperature
model of the extended emission is not self-consistent. Cooling takes
place for any reasonable choice of temperature of the X-ray gas.

\begin{figure*}
\begin{center}
\leavevmode
\epsfxsize 11 cm
\epsfbox{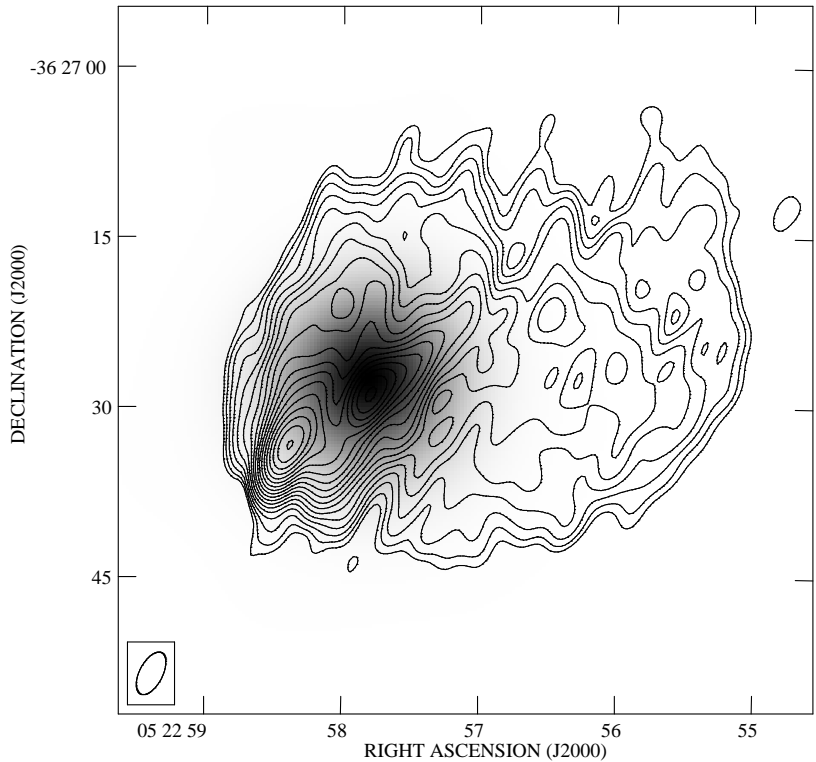}
\caption{1.4-GHz radio map of \source\ overlaid on a smoothed X-ray
image. The radio map was made from a 10-minute snapshot observation
with the VLA in BnA configuration, taken on 1989 Feb 14. The Gaussian
restoring beam is $4.09 \times 2.03$ arcsec, as shown in the bottom
left of the figure; contours are at $8 \times (1, \protect\sqrt 2, 2,
2\protect\sqrt 2, \dots)$ mJy beam$^{-1}$). Note the hot spot to the
SE and the jet extending NW from the core. The (dewobbled) X-ray
image has been smoothed with a Gaussian with $\sigma = 4$ arcsec;
black corresponds to 3.8 counts arcsec$^{-2}$.}
\label{overlay}
\end{center}
\end{figure*}

The extended X-ray component detected by the HRI is comparable in size
to the extended radio emission in \source\, which might be thought to
suggest an explanation in terms of inverse-Compton (IC) scattering by
the energetic electrons in the radio lobes of photons from the cosmic
microwave background radiation (CMB) and from the BL Lac nucleus
itself (cf.\ Brunetti, Setti \& Comastri 1997). Fig. \ref{overlay}
shows the X-ray image superposed on a 1.4-GHz VLA radio map made from
data supplied by G.G.\ Pooley. Since the flux on the shortest
baselines of this radio observation (made at BnA configuration) is
equal to the single-dish Parkes radio flux (Wright \& Otrupcek 1990)
of 16.3 Jy, we can be confident that the extended structure is well
represented. The largest angular size (LAS) of the radio emission is
about 50 arcsec, whereas the FWHM of the extended X-ray component is
10 arcsec, and Fig.\ \ref{overlay} shows that the radio emission is
quite asymmetrically distributed around the nucleus, in contrast to
the symmetry of the X-ray emission. However, we note that
inverse-Compton emission would come preferentially from the parts of
the radio lobes closest to the active nucleus, so this in itself does
not make an inverse-Compton model implausible.  Since we do not know
the unprojected geometry of the source, we model the extended
radio-emitting component crudely as a uniform sphere centred around
the nucleus. The flux density from the core is 2.7 Jy [consistent with
the value of Antonucci \& Ulvestad (1985)] leaving 13.6 Jy in extended
emission. Projection effects make it difficult to assess the true size
(and therefore volume) of the lobes, but the flux expected from IC
scattering from an object of a given synchrotron (radio) flux is only
weakly dependent on the object's size, and so we assume a radius for
the sphere of half the LAS. Because the frequency of a scattered
photon is raised by approximately the square of the Lorentz factor
($\gamma$) of the scattering electron, and because the electron energy
spectrum is dominated by electrons with $\gamma \sim 100$
[i.e. electrons near the assumed low-energy cutoff in the energy
distribution; cf. Carilli \etal\ (1991)] the most significant
contribution in the X-ray band from the IC process is due to far-IR
photons. We accordingly model the spectrum of the BL Lac's nucleus in
this region as a power law with spectral index 0.78 normalised to the
measured 100-$\mu$m flux of Impey \& Negebauer (1988). Pian \etal\
(1996) argue that the Doppler factor in the nucleus is of order unity,
based on the relatively low radio core prominence discussed in section
\ref{unbeamed} and on models for the nuclear X-ray and $\gamma$-ray
emission, and we adopt this value in order to infer a luminosity from
the observed flux; we neglect the effects of possible circumnuclear
obscuration, so that our source model is effectively an isotropically
radiating central source surrounded by a uniform, spherically
symmetrical electron distribution. Using a modified version of the
code of Hardcastle, Birkinshaw \& Worrall (1998a) we find that the IC
process in this model cannot produce enough X-rays to account for the
observed flux unless the lobes have magnetic field strength
approximately an order of magnitude weaker than the value implied by
equipartition of energy between magnetic field and relativistic
electrons (for electron filling factor unity); this is relatively much
weaker than fields that have been inferred in other sources from
observations of X-ray emission attributed to inverse-Compton or
synchrotron-self-Compton emission (Harris \etal\ 1994; Feigelson
\etal\ 1995; Hardcastle \etal\ 1998a; Brunetti \etal\ 1998).  At
equipartition, the X-rays from the IC process contribute at most 2 per
cent of the observed flux, or less if the filling factor is lower or
there is a significant contribution from relativistic protons.

As a result of this simple analysis, we believe that it is unlikely
that the extended X-rays seen around the source are due in large part
to inverse-Compton emission. Instead, it seems more likely that they
are indeed thermal emission from a rapidly cooling central region, and
that \source\ inhabits a cooling flow; this means that its environment
is significantly different from those of the FRIs observed by Worrall
\& Birkinshaw (1994), which tended to lie in less dense environments
with much longer cooling times. This model is qualitatively consistent
with the bright extended emission line region seen in \source\
(Boisson, Cayatte \& Sol 1989) and with the strong polarization
asymmetry seen in the radio observations.

We attempt to fit the radial profile of the source with a cooling flow
model based on that described by Hardcastle, Lawrence \& Worrall
(1998b). In this version of the model, the outer regions of the source
are fit with an isothermal $\beta$ model while the temperature and
density inside the cooling radius ($r_{\rm cool}$) are power law
functions of radius. The electron density follows
\[
n_e(r) = 
\cases{
n_{\rm e1}
   &$r \le r_{\rm inner}$\cr
n_{\rm e2} \left({r\over {r_{\rm cool}}}\right)^{-a}
   &$r_{\rm inner} \le r \le r_{\rm cool}$\cr
n_{\rm e3}\left(1+{r^2 \over r^2_{\rm core}}\right)^{-{3\over2}\beta}
   &$r_{\rm cool} \le r$\cr
}
\]

while the temperature is given by
\[
T(r)=
\cases{
T_{\rm e1} 
   &$r \le r_{\rm inner}$\cr
T_{\rm e2} \left({r\over {r_{\rm cool}}}\right)^{-b}
   &$r_{\rm inner} \le r \le r_{\rm cool}$\cr
T_{\rm e3}
   &$r_{\rm cool} \le r$\cr
}
\]

Here $r_{\rm core}$ is the core radius of the $\beta$ model, and
$r_{\rm inner}$ is the inner limiting radius of the cooling flow,
allowing us to avoid infinities at $r=0$. $n_{\rm e1}$, $n_{\rm e2}$
and $n_{\rm e3}$ are scale electron densities and $T_{\rm e1}$,
$T_{\rm e2}$ and $T_{\rm e3}$ scale temperatures; $T_{\rm e3}$
corresponds to the temperature of the non-cooling gas. It is clear
that matching temperatures and densities allows us to write all the
scale factors in terms of $n_{\rm e3}$ and $T_{\rm e3}$. The results
are only weakly dependent on $r_{\rm inner}$ so long as it is small
and we fix it at a value corresponding to 0.01 arcsec in what follows.

The parameters $a$ and $b$ set the slope of the density and
temperature distributions; for a realistic cooling model we expect
$a>0$ and $b<0$, so that density increases and temperature decreases
with decreasing radius. The ideal gas law implies that pressure goes
as $r^{-(a+b)}$. If the atmosphere is required to be close to
hydrostatic equilibrium, then matching mass as a function of radius
inside and outside the cooling zone leads to a constraint on $r_{\rm
cool}$ as a function of $a$ and $b$:
\[
{r_{\rm cool}\over r_{\rm core}} =
\left({{a+b}\over{3\beta-(a+b)}}\right)^{1/2} \]
and if we assume some
law for the radial dependence of pressure ($P \propto r^{-1}$ is
consistent with observations of other cooling regions) then the model
has only five free parameters: $\beta$, $r_{\rm core}$, $a$, $T_{\rm
e3}$ and $n_{\rm e3}$, the last being a normalisation parameter that
can be determined in the fit. In addition, such a model can only be
physically realistic if the cooling time at $r_{\rm cool}$ is comparable to the system lifetime, or the Hubble time:
\[
\tau_{\rm cool} = {{30 \times 10^6 \sqrt{k_B T}}\over{n_p(r_{\rm
cool})}} \la 2 \times 10^{10} {\rm\ years}
\]
(Sarazin 1986) where $\tau_{\rm cool}$ is in years, $k_B T$ in keV and
$n_p$ in cm$^{-3}$.

We fit a range of representative cooling models to the data. As
before, $\beta$ was chosen from a small number of possible values
(0.5, 0.667, 0.9) while $r_{\rm core}$ ranged from 1--1000 arcsec. We
tried values of 1, 1.5, 2, 5 and 10 keV for $k_B T_{\rm e3}$. $a$ was
allowed to take the values 1.5, 1.75 or 2.0. No model consisting of a
cooling flow alone was a good fit to the data. A number of models
consisting of a cooling flow and a central point source were
comparable in goodness of fit to the combination of $\beta$-model and
point source discussed above, but most of these were ruled out by the
constraint on cooling time. Those that were not have small core radii,
low external temperatures and steep temperature and density power
laws: for example, the model with $k_B T_{\rm e3} = 1.5$ keV, $a=2.0$,
$\beta=0.667$, $r_{\rm core} = r_{\rm cool} = 40$ arcsec is an
acceptable fit to the HRI radial profile ($\chi^2=25$). In this model
just under 10 per cent of the total counts ($1000 \pm 100$) are
assigned to the cooling flow component. The nominal cooling time at
the core radius is $2 \times 10^{10}$ years; the mass deposition rate
is approximately $40 M_\odot$ yr$^{-1}$ and the implied densities at
20 kpc correspond roughly with those required for pressure equilibrium
with the cold gas inferred from observations of the extended
emission-line region (Boisson \etal\ 1989). Though the details of the
model may not be correct, this shows that the extended emission of
\source\ can plausibly be modelled as a cooling flow of this type.  A
strong nuclear source is still necessary; the flux of the point-like
component in the cooling-flow model is reduced by only $\sim 1$ per
cent compared to that derived from the simple $\beta$-model fits of
section \ref{analysis}.

\section{Conclusions}

We have carried out the first separation of nuclear and galaxy-scale
extended X-ray emission in a BL Lac object and found evidence that
\source\ inhabits a dense and, presumably, rapidly cooling region of
X-ray emission, a much more extreme environment than those found for
`typical' FRI radio galaxies; we have shown that it can be modelled as
a cooling flow in low-temperature cluster gas. If this result were
extended to other BL Lac objects, it would cause difficulties for
models that seek to unify FRIs and BL Lacs, and might imply some
causal relationship between a dense and rapidly cooling atmosphere and
the BL Lac phenomenon. However, it may be that \source , with its high
power and intermediate radio structure, is not representative of the
BL Lac class. Further observations are planned both to verify the
thermal nature of the galaxy-scale halo and to see whether \source\ is
unusual among BL Lacs in this respect.

\section*{Acknowledgements}

We are grateful to Guy Pooley for allowing us to use his VLA
observations of \source , and to the VLA Analysts for help in
recovering the VLA data from the archive. This research has made use
of the NASA/IPAC Extragalactic Database (NED) which is operated by the
Jet Propulsion Laboratory, California Institute of Technology, under
contract with NASA. The Digitized Sky Surveys were produced at the
Space Telescope Science Institute under U.S. Government grant NAG
W-2166. The National Radio Astronomy Observatory is operated by
Associated Universities Inc., under co-operative agreement with the
National Science Foundation.  This work was supported by NASA grant
NAG 5-2312 and PPARC grant GR/K98582.

\clearpage

\end{document}